# Environmental Economics and Uncertainty: Review and a Machine Learning Outlook

Ruda Zhang, Patrick Wingo, Rodrigo Duran, Kelly Rose, Jennifer Bauer, Roger Ghanem

**Summary**

Economic assessment in environmental science concerns the measurement or valuation of environmental impacts, adaptation, and vulnerability. Integrated assessment modeling is a unifying framework of environmental economics, which attempts to combine key elements of physical, ecological, and socioeconomic systems. Uncertainty characterization in integrated assessment varies by component models: uncertainties associated with mechanistic physical models are often assessed with an ensemble of simulations or Monte Carlo sampling, while uncertainties associated with impact models are evaluated by conjecture or econometric analysis.

Manifold sampling is a machine learning technique that constructs a joint probability model of all relevant variables which may be concentrated on a low-dimensional geometric structure. Compared with traditional density estimation methods, manifold sampling is more efficient especially when the data is generated by a few latent variables. The manifold-constrained joint probability model helps answer policy-making questions from prediction, to response, and prevention. Manifold sampling is applied to assess risk of offshore drilling in the Gulf of Mexico.

## 1. Introduction

Earth system sustainability has been challenged across multiple planetary boundaries, such as climate change and biosphere integrity (Steffen et al., 2015). Environmental economics is concerned with such issues of global sustainable development, where a major topic is the valuation or measurement of environmental impacts.

Environmental valuation research covers events on varying time scales: from oil spills (Boyle & Parmeter, 2017), to invasive species (Eiswerth, Lawley, & Taylor, 2018), to climate change (Markandya, Paglialunga, Costantini, & Sforna, 2017; Smith, 2017). And it often involves valuation of ecosystems, such as coastal and marine ecosystems (Vassilopoulos & Koundouri, 2017), including mangroves (Barbier, 2017). However, economic assessment in environmental science is often subject to limited or costly data. Thus, it is important to properly characterize the uncertainty associated with impact assessment. This article reviews the issue of uncertainty in environmental impact assessment, focusing on the techniques of uncertainty characterization.

A machine learning technique is applied to characterize uncertainty in impact assessment, which constructs a joint probability model that helps answer three types of policy-making questions: (1) prediction; (2) response; and (3) prevention.

This article continues as follows. Section Two reviews the integrated assessment framework, a unifying framework of environmental economics, which provides the context. Section Three reviews the treatment of uncertainty in integrated assessment, and introduces manifold sampling, a new technique for uncertainty characterization. Section Four is a case study in environmental economic assessment, where manifold sampling is applied in an integrated assessment of offshore deepwater oil spills. Policy-making applications of the joint probability model built from manifold sampling is discussed in Section Five. This article concludes with possible improvements and future directions.

## 2. The Integrated Framework of Environmental Economics

Environmental economics is an interdisciplinary field connecting the natural, social, and statistical sciences. Despite its diverse research topics, a unifying framework has emerged from environmental economics. Integrated assessment modeling (IAM) attempts to combine key elements of biophysical and socioeconomic systems, which is illustrated in fig. 1.

A fully integrated assessment involves variables from all three component systems, each with models connecting variables internal to the system, while the gaps among systems are closed by models assessing impact, adaptation, and vulnerability. The component models are connected in a causal loop: socioeconomic systems provide environmental drivers to physical systems and ecosystems, which in turn causes environmental and ecological changes that impact the functioning of socioeconomic systems. An integrated assessment model does not have to include all three systems, or all components of any system. The decision for inclusion depends on the research question, the set of relevant variables, and the knowledge and resources accessible to the modeler. As a result, this framework allows for different configurations of variables and linkages.

In climate change research, for example, economic models of the human society determine carbon emissions and other factors that affect radiative forcing on Earth's atmosphere, based on which physical models of the climate system project future climate scenarios, which in turn translates to social outcomes using impact models. The numerous integrated assessment models (IAMs) of climate change may be partitioned into two groups, based on the types of components included: policy evaluation models (PEMs), or process-based IAMs; and policy optimization models (POMs), or benefit-cost IAMs (Farmer, Hepburn, Mealy, & Teytelboym, 2015; Houser et al., 2015). Policy evaluation models connect social system unidirectionally to physical system, and are used to assess the cost-effectiveness of achieving a particular mitigation target with a given policy. Policy optimization models (POMs) complete the feedback loop with impact models, and are used to find the optimal policy via cost-benefit analysis, weighing off the damage and the mitigation cost of climate change.

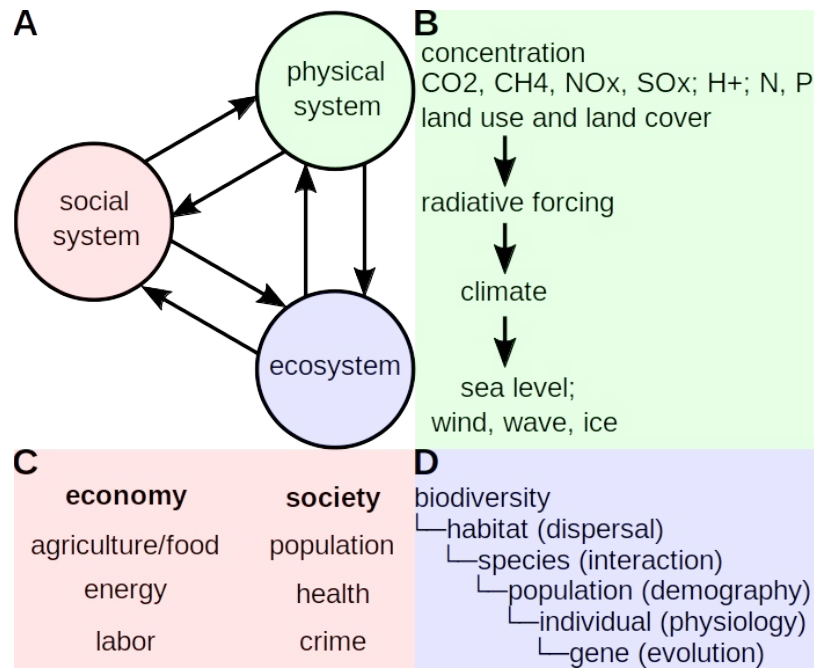

*Figure 1: The integrated assessment framework. System diagrams of: (A) a general integrated assessment model, and (B-D) physical, social and ecological systems. Arrows represent models linking system components. Credit: the authors.*

IAMs are also used for various other issues in environmental economics. For example, in coastal flood management, models for storms, floods, damages, and protections are integrated to aid resilience planning and investments (Aerts et al., 2014). For regional policy-making relevant to climate mitigation and adaptation, Kraucunas et al. (2015) integrates models of climate, hydrology, agriculture and land use, energy, and socioeconomic systems. IAMs that integrate climate models, food systems, and health models have been used to study the effect of climate change via food production on mortality (Springmann et al., 2016). IAMs for biodiversity and ecosystem services are also been built (Pereira et al., 2010; Dawson, Jackson, House, Prentice, & Mace, 2011), with ongoing coordination efforts resembling that of the Intergovernmental Panel on Climate Change (Díaz et al., 2015) and for building mechanistic ecosystem models (Urban et al., 2016).

To have a closer look at the three components of IAM, again, take climate change as an example.

Climate models represent the Earth's natural systems to study how climate responds to natural and human-induced perturbations, which may be built at different complexity levels (Moss et al., 2010). Global climate models (GCM), also known as atmosphere–ocean general circulation models (AOGCM), simulate the dynamical and physical processes of atmosphere, ocean, land, river, and cryosphere, spatially resolved on grid cells. Regional climate models (RCM) are climate models on regional scales. Earth system models (ESM) are GCMs with chemical and ecological processes, such as land and ocean carbon cycle, atmospheric chemistry, aerosols, and vegetation.

Economic models represent key features of human systems to estimate future emission path and the cost of mitigation. In these models, a representative agent decides on consumption and investment in an inter-temporal optimization, subject to different market equilibrium assumptions (Farmer et al., 2015). Partial equilibrium models optimize a specific sector of the economy such as energy supply, and consider investment in clean and dirty energy technologies. General equilibrium models, on the other hand, optimize the whole economy. For example, computable general equilibrium (CGE) models link different countries and different economic sectors in one country through trade, and may use sectoral production functions, inter-sectoral spillovers, and international trade relationships.

Impact, adaptation, and vulnerability assessment adopts a wide range of approaches. Following standard definitions of these terminologies (O'Neill et al., 2017): *impact* is the manifestation of risk; *hazard* is a physical event, trend, or impact that may cause ecological, social, or economic impacts; *exposure* is the presence of ecosystems, people, or assets under hazards; *vulnerability* is the susceptibility and predisposition to harm of the exposed entities; and *adaptation* is adjustment to actual or expected hazard and its effects.

Here the focus is on impact assessment, which relate climate events (such as temperature, precipitation, and extreme events) to social and economic outcomes. Approaches to impact assessment can be broadly categorized into valuation or enumerative methods (Tol, 2009), and statistical or econometric methods (Hsiang, 2016). Impact models with no empirical basis are excluded, which is the case for damage functions in traditional narrowly-defined IAMs of climate change such as DICE (Nordhaus, 1992).

Enumerative methods collect published estimates of the physical impacts of climate change based on laboratory experiments, climate models, or impact models, give a price to each physical impact and add them up for the total economic cost. Valuation of physical impacts can be done using market price of goods, cost of mitigation measures, or by benefit transfer for nonmarket goods and services such as ecosystem services. Benefit transfer is the adaptation of existing value estimates for another valuation study for a different time or location (Boyle & Parmeter, 2017). Benefit transfer may be categorized into three types: value transfer uses the value estimate from one study site, or the average from multiple study sites; function transfer uses an econometric model estimated from one study site; meta-function transfer uses an econometric model estimated from multiple study sites, optionally imposing a structure of preferences.

Econometric methods apply regression models of social and economic outcomes on climate variables, using observed variations over space (Mendelsohn, Nordhaus, & Shaw, 1994) and time (Deschênes & Greenstone, 2007). Social and economic outcomes may be measured by sector and context, such as agriculture, energy demand, trade, labor, health, demographics, and conflict (Carleton & Hsiang, 2016). Total economic outcome may be measured in aggregate accounts such as GDP ("top-down" approach) (Burke et al., 2015b), or as the sum of outputs in different economic sectors ("bottom-up" approach) (Tol, 2002; Hsiang et al., 2017).

## 3. Uncertainty in Integrated Assessment

The integrated assessment models in environmental economics have been criticized for their many shortcomings (Ackerman, DeCanio, Howarth, & Sheeran, 2009; Pindyck, 2013; Stern, 2016). A major criticism is their treatment of uncertainty. The argument is straightforward: if environmental economic research is to inform the public about the potential future effects of environmental change and mitigation policy, researchers must communicate the uncertainties associated with these estimates, and properly model human decision-making under uncertainty.

Uncertainty affects environmental policy evaluation. Classic decision theory models human preferences with a utility function: if the outcome of a policy is uncertain, each policy is valued by the expected utility of the uncertain outcomes. Utility function captures the decision maker's attitude towards risk, e.g. risk aversion, which implies a negative effect of uncertainty on policy value. Even without assuming risk aversion, similar effect of uncertainty exists if environmental change is irreversible (Arrow & Fisher, 1974). Because the standard expected utility theory has other limitations such as the restrictive assumptions from the von Neumann–Morgenstern utility theorem (Neumann & Morgenstern, 1947), alternative decision-making frameworks have been advocated, such as robust decision-making tools (Hall et al., 2012) and non-probabilistic approaches (Heal & Millner, 2014).

Various issues in integrated assessment have been identified as uncertainty (Heal & Kriström, 2002; Asselt & Rotmans, 2002; Pindyck, 2007). It is insightful to categorize these "uncertainties" into three types: (1) discrepancy between model and reality; (2) model sensitivity; and (3) data variation. Discrepancy from reality exists for all models of physical, ecological, and socioeconomic systems. For example, climate models lack comprehensive representation of Earth processes such as carbon cycle, which affects climate sensitivity estimates; and the discrete representation of Earth surface necessarily sacrifices variances within each spatial unit (Moss et al., 2010; Stern, 2013). Economic models adopt parameters that are difficult to calibrate, such as discount rate and rate of technological change (Heal & Kriström, 2002), which affects emissions pathway estimates. Economic models also rarely capture microeconomic adaptation pathways and general equilibrium effects such as price adjustment and factor reallocation, which may significantly alter climate impacts (Kahn, 2016; Barreca, Clay, Deschenes, Greenstone, & Shapiro, 2016). Model sensitivity to the choice of model form and parameters is evident in climate models and particularly so in impact models. Traditional IAMs adopt arbitrary, highly nonlinear function forms for the damage function, and set discount rates relatively high for very long-time horizons (Pindyck, 2007; Ackerman et al., 2009). Data variation refers to the differences within a data set, either from empirical observation or model simulation, such as weather station data and output of climate model experiments. This last category is the focus and adopted notion of uncertainty in this article.

Uncertainty, or data variation, have been discussed in literature from different aspects and in different numbers of dimensions. For a single variable, the presence of variation itself already implies adjustments to expectation-based results (Arrow & Fisher, 1974). Much attention has been given to the tail of the probability distribution, i.e. low-probability high-impact events (Weitzman, 2009; Pindyck, 2013; Stern, 2013; Burke et al., 2015a).

For two variables related by nonlinear transformation, the probability distribution of the response variable can be much distorted from that of the input. This explains the inherent uncertainties about climate sensitivity (Roe & Baker, 2007) and tipping points in physical and economic impacts, e.g. permafrost thawing, methane release, human migration and the resulting conflict (Lenton et al., 2008). Multiple variables may exhibit correlation structure, which may lead to very different results than if assumed to be mutually independent. The physical and socioeconomic impacts of environmental change are generally treated independently in the literature, ignoring their interactions, which may severely underestimate the total cost (Cai, Lenton, & Lontzek, 2016).

### *3.1 Characterizing Uncertainty*

Many researchers consider it crucial to explicitly incorporate uncertainty in integrated assessment (Kopp, Hsiang, & Oppenheimer, 2013; Farmer et al., 2015; Burke et al., 2015a).

Characterizing and propagating uncertainty in physical models, and economic equilibrium models, can use well-established methods (Ghanem & Spanos, 1991; Soize & Ghanem, 2004). In climate change studies, an ensemble of simulations is often used to assess uncertainty: multi-model ensembles collect and assign equal weights to model outputs from the participating modeling centers (Taylor, Stouffer, & Meehl, 2012; Kay et al., 2015); perturbed physics ensembles use a single physical model and set model parameters to combinations of plausible values (Stainforth et al., 2005). Perturbed ensembles have also been applied to economic equilibrium models and traditional IAMs (Gillingham et al., 2018). Note that neither type of ensemble represents the underlying probability distribution of the outcome: multi-model ensembles collect best guesses of the participating modelers; while perturbed ensembles are typically factorial experiments on the parameter space. Monte Carlo sampling methods can construct a probability distribution on the outcome, given probability distributions of the input. But because repeated simulations of complex physical models are computationally demanding, simplified models and model surrogates may be used to apply this method (Rasmussen, Meinshausen, & Kopp, 2016).

Characterizing uncertainty in impact models, on the other hand, requires different approaches. This is because currently there lacks a comprehensive understanding about the pathways connecting physical and socioeconomic systems, and thus do not have impact models that are mechanistic. In fact, current impact models either are deterministic or focus only on expected values of random variables.

Damage functions in traditional IAMs take conjectured function forms, typically convex, e.g. quadratic functions, relating global mean temperature change to aggregate loss (Farmer et al., 2015). Adding uncertainty into this type of impact models would not improve their credibility, as they lack empirical foundation to start with (Pindyck, 2017).

Enumerative methods take speculative monetary valuation of climate impacts on ecosystems and human health and life, often with income bias, i.e. lives of citizens of rich countries are valued much more than those of poor countries (Ackerman et al., 2009). If obtained by benefit transfer, such valuation is further subject to the lack of ecological and economic correspondence between the locations and across time (Plummer, 2009). These limitations prevent incorporation of uncertainty from being helpful.

Econometric methods take a more rigorous empirical approach to impact assessment (Hsiang, 2016). The goal to is estimate some dose-response function (Carleton & Hsiang, 2016), a regression function of a socioeconomic outcome on environmental and non-environmental variables. For example, a panel model for short-run (e.g. annual) weather effects can be specified as (Dell, Jones, & Olken, 2014):

$$y_{it} = \beta C_{it} + \gamma Z_{it} + \mu_i + \theta_{rt} + \varepsilon_{it} \tag{1}$$

Here, subscripts $i$ and $t$ index geographic units and time, variable $y$ denotes a socioeconomic outcome, $C$ and $Z$ denote weather variables and non-climatic controls, $\mu$ and $\theta$ denote spatial fixed effect (unit-specific benchmark level) and time fixed effect (time trend), and $\varepsilon$ is the error term, or unexplained variations. In general, such econometric models are not necessarily linear in the weather variables, or homoscedastic in the error terms (variance independent of the regressors). But econometric models also have drawbacks, perhaps the most critical of which is model specification. Common specifications of an econometric model involves assumptions that are not falsifiable: the set of regressors to include; that the outcome is a separable function of the regressors; and the correlation structure in the error terms.

With explicit uncertainty characterization in all components of an integrated assessment, uncertainty of the overall impact can be constructed. Hsiang et al. (2017) constructed probability distributions of climate scenarios using Monte Carlo methods and mapped them to sectoral damages using dose-response functions—which are conditional probability distributions—to derive probability distributions of socioeconomic impacts. These probabilistic impact estimates are also supplied to an economic model to adjust for general equilibrium effects. To calculate the aggregate impact on GDP, however, they used enumerative valuation of sectoral damages, including the value of a statistical life which dominates when warming exceeds 3 °C.

*3.2 Manifold Sampling*

An alternative approach to characterize uncertainty in integrated assessment is manifold sampling. Manifold sampling (Soize & Ghanem, 2016) is a machine learning technique, which is model-free (or non-parametric), constructs a joint probability model of all relevant variables, and suitable for high dimensional data. Such method is especially attractive for environmental science problems, where the dimensionality of physical variables and socioeconomic impacts is often greater than ten.

Manifold sampling is a technique to sample from a joint probability density that is estimated from an original sample. Unlike traditional kernel density estimation, manifold sampling generates samples that are concentrated on the geometric structure, i.e. manifold, defined by the original data. When data are assumed to conform to certain geometric structure such as a response surface, variations in possible new observations should also be subject to such constraint. Disregarding this geometric structure can sacrifice the effectiveness of non-parametric probability density estimates, which is particularly severe for sparse data in high-dimensional spaces, which is the case for environmental science problems. Manifold sampling improves traditional density estimates by projecting them onto a reduced-order diffusion maps basis, which preserves main features of the manifold learned from data.

The procedure of manifold sampling is detailed in Soize & Ghanem (2016), and only a brief overview is provided here. Suppose one is interested in the probabilistic relationship among a set of variables $X$, which is modeled by a joint probability distribution $p_X$ such that $X \sim p_X$. The available observations on random vector $X$ is presented as a sample matrix $[x]$. First one can build a probability density estimate $p_H$ using $[x]$, for example, by Gaussian kernel density estimation, so that $p_H$ approximates $p_X$. Separately, characterize the manifold learned from data $[x]$ using a diffusion maps basis $[g]$, which is truncated to the first few vectors $[g_m]$ for a reduced-order representation of the manifold. Now one can generate a new sample matrix $[h]$ from $p_H$ using Monte Carlo methods, and then project it onto $[g_m]$ to keep them concentrated around the manifold. More sample matrices can be generated by repeating the last step. Sections 4 and 5 show how to apply this technique to characterize uncertainty in integrated assessment and use its result to inform policy decisions.

## 4. Case Study: Offshore Oil Spills

One environmental problem of severe consequences are offshore oil spills (Nelson, Bauer, & Rose, 2014). Unlike oil spills in contained water bodies such as harbors, rivers, and lakes, these events occur away from the coast and potentially can cause greater harm to the environment, economy, and society.

Offshore oil spills have impacted and been an ongoing risk to global waters since the onset of offshore drilling and transport activities worldwide, see table. 1. Of the world's most significant offshore spill events, the 2010 Deepwater Horizon (DWH) blowout in the Gulf of Mexico was unprecedented both in scale and location. This spill resulted from a loss of control event associated with the Macondo well, a deepwater oil well with an ocean depth of over 1.5 kilometers. Deepwater, and ultra-deepwater, refer to water depths greater than 1,000 ft / 5,000 ft (305 m / 1,524 m) (Nixon, Kazanis, & Alonso, 2016), or 500 m / 1,500 m (Cummings et al., 2014). Whereas shallow water wells locate on the continental shelf, deepwater wells are typically on the continental slope, where water depth poses a greater challenge to petroleum production technology, as well as response strategy in case of failure. The DWH blowout ultimately led to a release of 3.19 million barrels of oil and 1.84 million gallons of dispersant into the Gulf of Mexico (Deepwater Horizon Natural Resource Damage Assessment Trustees, 2016). While the DWH well was the largest accidental spill volumetrically, even smaller spills, such as the Exxon Valdez spill offshore Gulf of Alaska (Nelson et al., 2014), or spills of longer duration but lower flow rates, such as the Taylor well (Sun et al., 2018), can result in cumulative and lasting impacts to the environment and economics of offshore regions.

*Table 1: Examples of Major Offshore Oil Spills (Volume in million barrels)*

| Start | End | Name | Cause | Location | Volume |
|---|---|---|---|---|---|
| 1979-06-03 | 1980-03-23 | Ixtoc I | blowout | Gulf of Mexico | ~3 |
| 1980-01-17 | 1980-01-30 | Funiwa No. 5 | blowout | Niger Delta | ~0.42 |
| 1991-01-23 | 1991-05 | Gulf War | dumping | Persian Gulf | 2–4 |
| 1991-04-11 | 1991-04-14 | MT Haven | explosion | Mediterranean Sea | ~1 |
| 2010-04-20 | 2010-07-15 | Deepwater Horizon | blowout | Gulf of Mexico | 3.19 |
| 2018-01-06 | 2018-01-14 | Sanchi | collision | East China Sea | 0.96 |

Damages caused by these offshore oil spills can be analyzed from two perspectives: one of the human society, and one of the ecosystem. Multiple socioeconomic ocean use sectors are disrupted by offshore oil spills, such as the oil and gas industry, commercial transportation, commercial fisheries, and recreation and tourism. On the other hand, marine organisms are exposed to oil, dispersant, and other response induced stressors, through direct contact with oil, ingestion of contaminated water and sediments, consumption of contaminated prey, and presence in altered habitat. Such exposure can cause a wide array of toxic effects including death, disease, reduced growth, impaired reproduction, and physiological impairments that reduce the fitness of organisms. Collectively, these injuries can affect ecological processes and ecosystem services. For the Deepwater Horizon blowout, apparent system-wide population crashes were not observed among the surveyed fish and water column invertebrate species; however, the death of large numbers of larval fish and invertebrates represents a substantial short duration loss to the water column food web (Deepwater Horizon Natural Resource Damage Assessment Trustees, 2016).

This section presents an integrated assessment of offshore deepwater blowouts, characterizing uncertainty via manifold sampling. Integrated assessment modeling have been applied to oil spills by various parties (Bauer et al., 2015; Romeo et al., 2015; BOEM, 2017). One may simulate random deepwater blowouts on the Gulf of Mexico Outer Continental Shelf, track the pathways of discharged oil in the environment, calculate exposure metrics of the ecosystem and coastal community and economy, and build a joint probability model useful for policy-making. The model framework is illustrated in fig. 2.

*4.1 Environmental Models*

Environmental models represent current knowledge about the state of the physical environment, which provide a basis for numerical experiments. For this oil spill study, these include ocean model, digital elevation model, and crude oil composition.

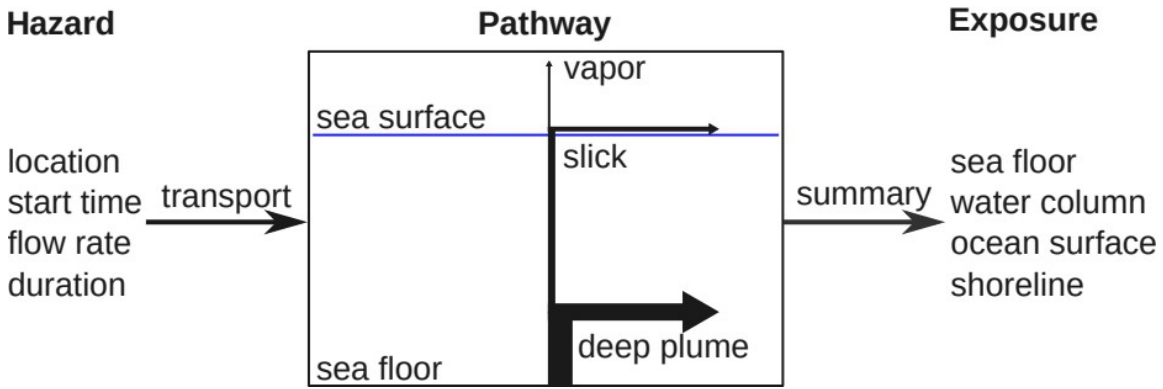

*Figure 2: An integrated assessment framework for offshore oil spills. Credit: the authors.*

Ocean models are representations of the ocean state discretized on a grid over time, often produced by assimilating observations with an ocean general circulation model (OGCM). Ocean state is a set of hydrodynamic variables, which may include ocean current velocity components, water temperature, water salinity, sea surface elevation, and horizontal wind stresses at the sea surface, which is a boundary condition. Regional ocean models covering the Gulf of Mexico includes the Hybrid Coordinate Ocean Model (HYCOM) Gulf of Mexico 1/25° Analysis (GOMl0.04), and the Navy Coastal Ocean Model (NCOM) American Seas (AmSeas). The NCOM AmSeas data have a higher horizontal resolution (1/30 degree grid, about 3 km in the Gulf of Mexico), include ocean surface wind stress, and align better with the coastlines. To save storage and computation time, the NCOM AmSeas data is spatially subset by a horizontal bounding box that extends south to west Cuba, east to Grand Bahama, and north and west to the Gulf Coast (see fig. 3).

Digital elevation models (DEM) represent the topography of the Earth's solid surface on a regular grid of geographic coordinates. Global bathymetric DEMs that are publicly available include: NOAA's ETOPO1 Global Relief Model, at 1 arc-minute resolution (Amante & Eakins, 2009); SRTM30_PLUS, at 30 arc-second resolution (Becker et al., 2009); and General Bathymetric Chart of the Oceans' GEBCO_2014, at 30 arc-second resolution (Weatherall et al., 2015). Here a subset of the GEBCO_2014 data is taken, with a spatial extent covering the ocean model subset, transformed into an Albers projection. Note that some ocean models, such as HYCOM, also provide the topographic models used in their numerical experiments.

Crude oil is a mixture of hydrocarbon components, which have different physical and chemical properties that affect their transport in the ocean. Crude oil properties by distillation cuts can be readily obtained from assay data, such as the NOAA's ADIOS OilLibrary, a publicly available library of oil chemistry data. Crude data from the Thunder Horse oil field (ADIOS ID 84010002) is used as a proxy for all deepwater oil fields, and assume that gases are well-mixed in bubbles.

### *4.2 Transport Model*

Oil spill models are physical models that simulate the transport of oil parcels in a body of water. For example, the General NOAA Operational Modeling Environment (GNOME) is a pollutant surface trajectory forecasting tool, which can be used to simulate oil spills from vessels (Beegle-Krause, 2001). Here the Blowout and Spill Occurrence Model (BLOSOM) is used, which supports four-dimensional fate and transport of oil parcels originated from the ocean bottom (Nelson, Grubesic, Sim, Rose, & Graham, 2015). Duran et al. (2018) compares these two oil spill models.

BLOSOM incorporates multiple physical mechanisms affecting oil transport, including ocean current advection, diffusion, and wind advection. Analytically, the governing equation of oil transport dynamics can be written as:

$$\frac{\partial P}{\partial t} + U \cdot \nabla P = \nabla \cdot (k \nabla P) + R \qquad (2)$$

This is the Fokker-Planck equation of the probability distribution of oil parcels, where $P$ is probability density, $t$ is time, $U$ is advection velocity, $k$ is diffusion coefficient, and $R$ is the source of released oil. The equation is analogous to the advection–diffusion equation, but uses an eddy diffusivity coefficient instead of molecular diffusivity. BLOSOM solves this hydrodynamic problem using the Lagrangian formulation, rather than the equivalent Eulerian formulation presented in Equation 2 (Sim et al. 2015; Duran, 2016).

Ocean models provide ocean current velocity and surface wind stress, but lack a small-scale "diffusive" velocity. The most common solution is to use large-scale ocean currents as an "advective velocity" and then add a stochastic component as the "diffusive velocity" (see for example Griffa, Piterbarg, & Özgökmen (2004)). BLOSOM also adopts this approach, setting the diffusion coefficient to account for the ocean model's resolution. A more complex alternative is not used: the random walk model with a spatially varying diffusion coefficient can cause agglomeration of particles in regions of small diffusivity (Duran, 2016).

BLOSOM also simulates weathering, the physical and chemical processes in the environment that alter crude oil properties. The weathering processes captured in BLOSOM includes evaporation (Mackay & Paterson, 1981; Jones, 1997), spreading (Fay, 1971; Lehr, Cekirge, Fraga, & Belen, 1984), emulsion (Fingas & Fieldhouse, 2004), dissolution (Riazi & Roomi, 2008), and water entrainment. BLOSOM currently does not model sedimentation, biodegradation, or photolysis.

BLOSOM takes a scenario file as input, which specifies all the environmental model input, blowout location, onset time, and duration, and other model settings. During the simulation, BLOSOM tracks the location and state of all the oil parcels, which can be recorded at regular time intervals.

### *4.3 Spill Scenarios*

Blowouts are simulated at the centroids of all the 152 currently productive deepwater blocks (Nixon et al., 2016, Appendix A; see also Fig. 3), with depths 50 meters above the ocean bottom inferred from the digital elevation model, to avoid oil sinking due to difference with BLOSOM's interpolation algorithm. The onset of blowouts may occur at different time of a year; specifically, one day is picked for every 30 days, 12 days in total, ending on 2018-04-02. In a climatological sense, this approach is an adequate approximation to oil spill variability due to blowout onset, because ocean circulation exhibits recurrent Lagrangian transport patterns, which can be captured in 12 climatological velocity fields per 30-day periods throughout a year (Duran, Beron-Vera, & Olascoaga, 2018, Gough et al. 2019). Thus, there are 1,824 distinct blowout scenarios, one for every combination of blowout location and onset date.

All the other model settings are controlled. Simulation time step is 10 minutes, blowout duration is 7 days, and simulation duration is 60 days. Longer simulation would not significantly affect the cumulative impact of oil spills, because slicks from a blowout lasting one week is likely to be almost entirely evaporated, or otherwise degraded, after a two-month period (Fingas, 1999). Crude flow rate is 50 barrels per day, about three orders of magnitude less than the estimated average hydrocarbon release rate of the Deepwater Horizon oil spill (Ryerson et al., 2012). This flow rate reflects a representative subset of released oil in a possible future blowout, and is limited by the number of oil parcels tracked in the simulation, which linearly increases time and space complexity. BLOSOM does not model wave advection, which may be responsible for the actual beaching of nearshore oil slicks through Lagrangian Stokes drift (Weisberg, Lianyuan, & Liu, 2017). Instead, wind advection coefficient is set to 0.03 and diffusion coefficient to 0.2 $m^2$/s. Note that Stokes drift may be simulated by applying wind directly (Clarke & Van Gorder, 2018). Response actions are not modeled; specifically, no dispersant treatment is applied throughout the simulation.

Results of the simulations are outlined in fig. 3, where the distribution of water column and surface oil are omitted for clarity. Note that although the deepwater oil fields are mostly located along the continental slope in the Northern Gulf, oil slicks may reach most parts of the ocean model boundary, except the west coast of Central and South Florida and the part over the Yucatan shelf. The difference in oil spill exposure across coastal regions can be explained by important kinematic features of surface ocean currents in the Gulf. These two exceptions to impacts by oil slicks are the coastlines isolated by wide, shallow continental shelves (Duran et al., 2018 and references therein). Although the Louisiana-Texas shelf is also wide, the northwest coastline of the Gulf is still subject to oil beaching, due to their proximity to oil fields, with many blocks just west of the Mississippi delta located within the persistent transport barriers. Coastal regions around the Mississippi River Delta are most likely to be affected by oil spills, not only because of their vicinity to the oil fields, but also due to coastal attractive currents. Oil spills originated from the Mississippi Canyon may also reach remote coasts such as north Cuba, Florida from the Keys to all its east coast, and the Bahamas. This can be attributed to the Loop Current, a region of persistent attraction which enters the Gulf from the Yucatan Channel between Mexico and Cuba and exits from the Florida Straits. Coasts of Mexico may also be affected due to meridional stretching (north-south advection) along the western Gulf; however, Mexican coasts along the Yucatan shelf are at relatively low risk. These results from simulations using instantaneous currents are also consistent with Duran et al (2018) that uses climatological currents. Note that the two studies use different numerical ocean models.

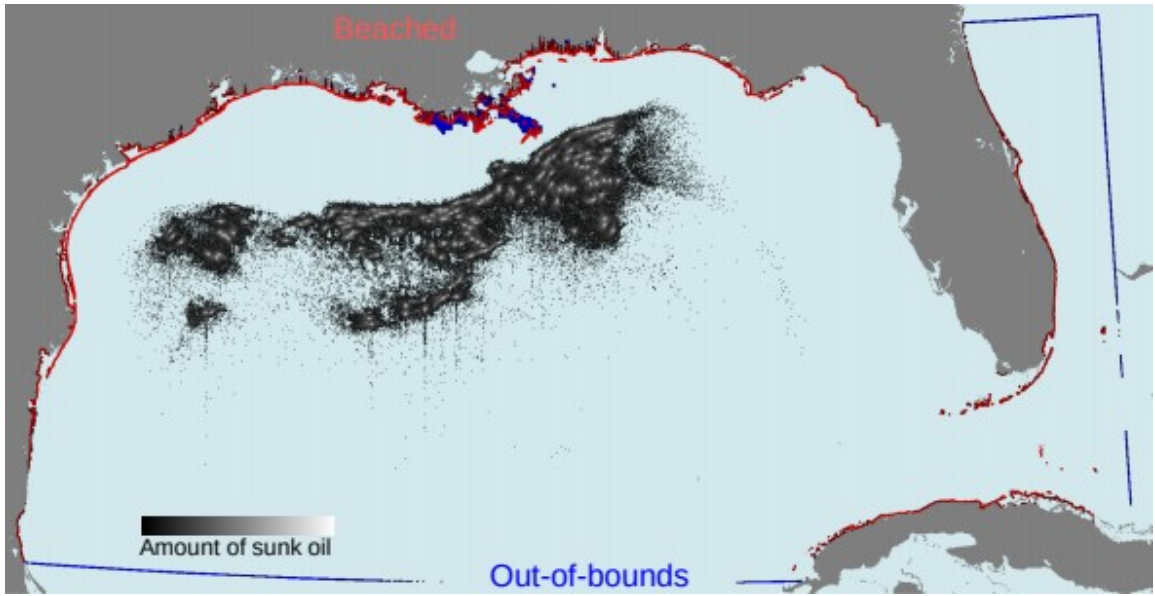

*Figure 3: Scope of simulation relative to the Gulf of Mexico. Beached (red) and out-of-bounds (blue) oil parcels from all simulations highlight the boundary of the ocean model. Areas with high amount of sunk oil (white) indicate deepwater oil fields. Credit: the authors.*

### 4.4 Exposure Metrics

Released oil can affect physical, ecological, and socioeconomic systems in various ways. To quantify the many dimensions of oil exposure, one needs to extract metrics from the oil spill scenarios. For the Deepwater Horizon oil spill, comprehensive assessment of its impact on ecosystem services and coastal economies has been carried out (Deepwater Horizon Natural Resource Damage Assessment Trustees, 2016, Chap. 4, see e.g. Tables 4.4-6, 4.5-3, 4.6-2). Following their analysis, several metrics are constructed for robust assessment of exposure.

Metrics of ecosystem exposure are categorized by the location of contact. Marine benthic ecosystems reside at the ocean bottom, whose exposure may be measured by the total area of sunk oil, sunkKM2, defined as the total square kilometers of sea floor ever in contact with released oil. Water column ecosystems are active in deep water and near ocean surface, which are subject to deep water oil plume, rising oil plume, surface slick, and subsurface entrained oil. Water column exposure is measured by two metrics: wcKGDay, the cumulative mass-day of water column oil, with unit kg-day; and slickKM2Day, the cumulative area-day of oil slick, with unit $km^2$-day. Nearshore ecosystems are distributed in estuarine coastal wetlands, beaches, and other habitats. Their exposure to beached oil is measured by beachKM, the total length of oiled coastline in kilometers, and beachKG, the total mass of beached oil in kilograms. These metrics may be refined by water depth, distance to shore, shoreline classification, species, and toxic concentration by contaminant group. This case study does not make such distinctions, and simply account for the cumulative exposure of major ecosystems.

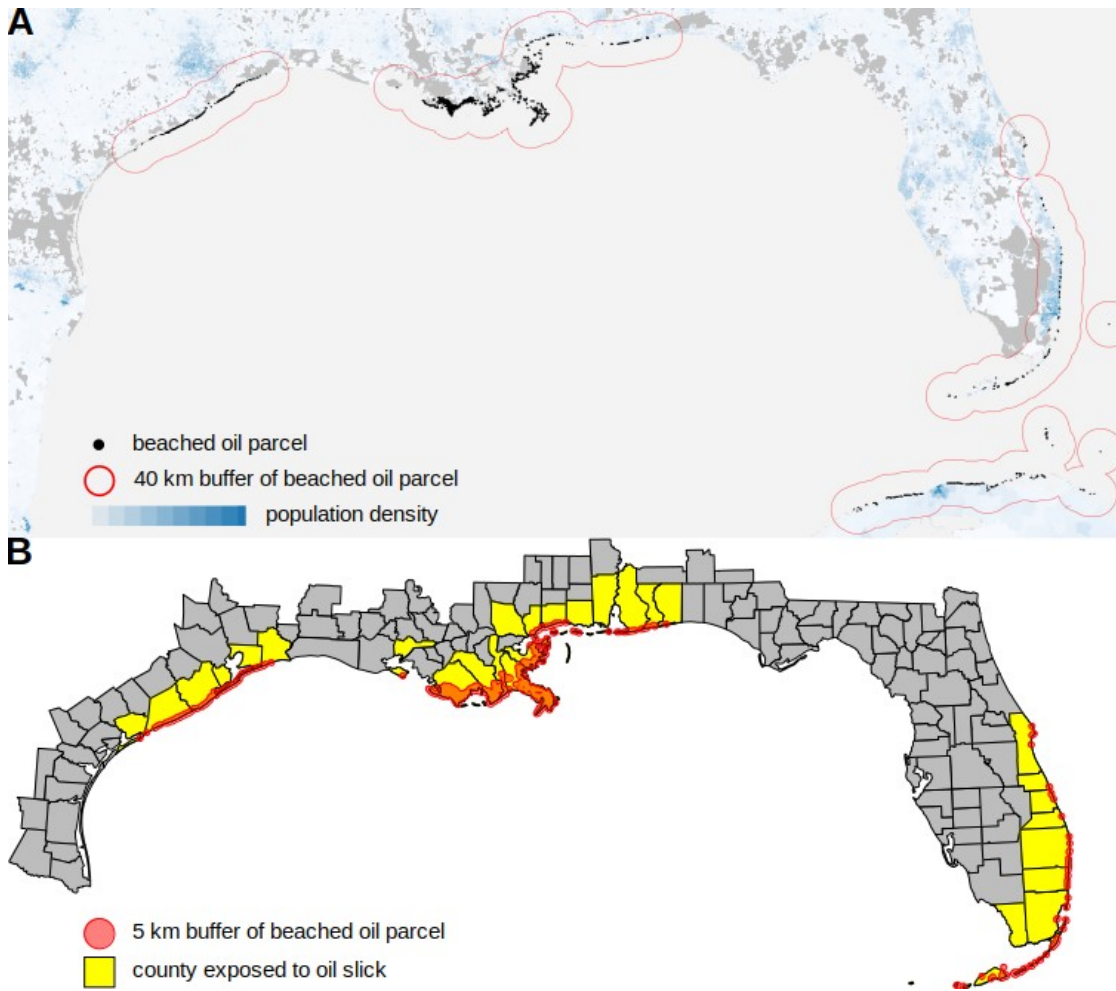

*Figure 4: Calculating socioeconomic exposure in (A) coastal population and (B) county-level economic activity. Oil spill scenario is the same for both panels. Cuba and the Bahamas are omitted in panel B due to a lack of county-equivalent level economic activity data. Credit: the authors.*

Metrics of socioeconomic exposure may be measured by the vicinity of coastal communities and economies to beached oil. Population exposure in each oil spill scenario is calculated by beachPop, the number of people within 40 kilometers of beached oil, and calculate economic exposure by salesKUSD, the total annual travel accommodation revenue of counties overlapping with the 5 kilometer buffer of beached oil, measured in thousand US dollars. The calculation is illustrated in fig. 4. Population data is from the Gridded Population of the World, which provides population density estimate for the year 2015 on a 30 arc-second grid (Center for International Earth Science Information Network (CIESIN) Columbia University, 2017). Economic activity data is from the 2012 Economic Census of the United States, which provides the latest annual county-level revenues from traveler accommodation (U.S. Census Bureau, 2012, Table EC1272A1).

## 4.5 Joint Probability Model

With data relating blowout scenarios and exposure metrics, a joint probability distribution that conform to the data manifold of these variables can be obtained via manifold sampling. Marginal distributions of the joint probability model are shown in fig. 5, although the high dimensional manifold is difficult to visualize.

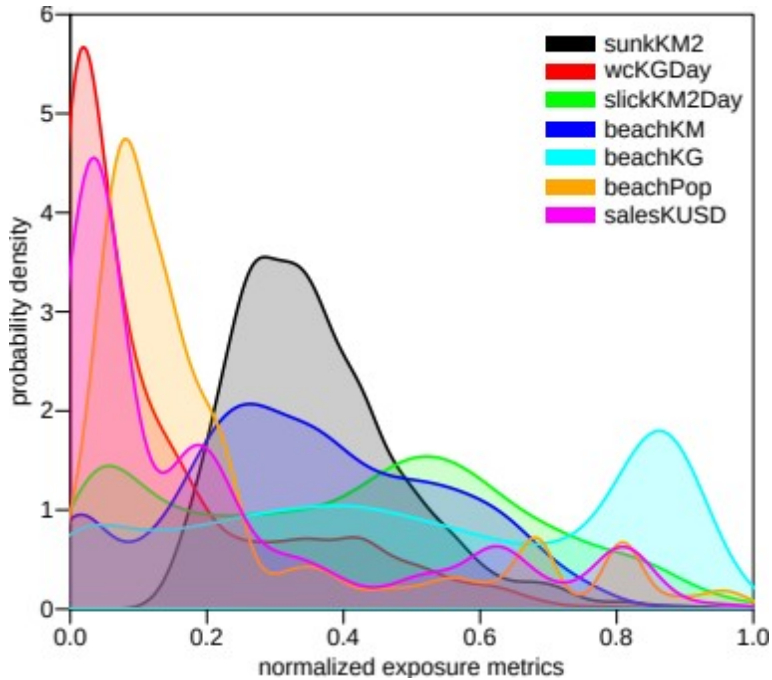

*Figure 5: Marginal distributions of the joint probability model of normalized exposure metrics, constructed via manifold sampling. Credit: the authors.*

The joint distribution estimated from the simulations can be interpreted as an uninformative prior, which has uniform marginal distributions over time and the currently productive deepwater oil blocks. The result can be enriched with extra simulation runs that take Monte Carlo samples of the ocean model from historical reanalyses and/or observations, to account for both climatological velocity and weather patterns. One may also extend the blowout duration for situations that take longer time to cap the blowout site. The prior distribution may be adjusted by weighting on blowout variables, such as time and location, to be consistent with historical observation and expert opinion. Note that although crude oil flow rate is hold constant in the simulations, which is only intended to be a representative fraction of a potential blowout, it should linearly affect spill density without changing spill extent.

## 5. Uses in Policy-making

Joint probability models such as those built from manifold sampling allow conditioning on different combinations of variables, unconstrained to the traditional inference of cause and effect of an impact model. Here discusses three types of policy questions that can be answered by such joint models, using deepwater oil spill as an example:

1. Prediction: Impacts of a hypothetical oil spill in the future.
2. Response: Real-time estimation of less observable quantities of an oil spill.
3. Prevention: Spatiotemporal risk distribution of deepwater drilling.

To simplify the discussion, let $X$ be the variables specifying a blowout scenario, $Y$ be the outcome variables of interest, and $P(X,Y)$ be the joint probability model.

The most straightforward question is to predict the impacts of a hypothetical future event. One may specify a blowout scenario $X$, and the conditional distribution $P(Y|X)$ is the answer. For example, one may ask what the environmental and economic impacts would be, if the Deepwater Horizon blowout happens in 2019. Or more generally, what if a deepwater oil platform in the Mississippi canyon blows out today? And, do the impacts depend on blowout depth and time? The last two questions are illustrated in fig. 6, where the two panels show $P(wcKGDay|z)$ and $P(beachKM|t)$ respectively. The cumulative mass-day of water column oil strongly depend on blowout depth, which increases quadratically in shallow and deep water, and increases linearly and sublinearly in ultra-deepwater. The total length of oiled coastline also varies by month, but with wider residual variation. In general, more coastal regions are subject to beached oil in March and April, while much less so in November and December. This observation can be largely attributed to recurring patterns in surface currents, and potentially wind direction: coastal attractive currents near the Mississippi River Delta are intense in spring. Additionally, the Loop Current may extend northmost in spring, while it will most likely have retracted south by the last three months of any given year (Duran et al., 2018 and references therein).

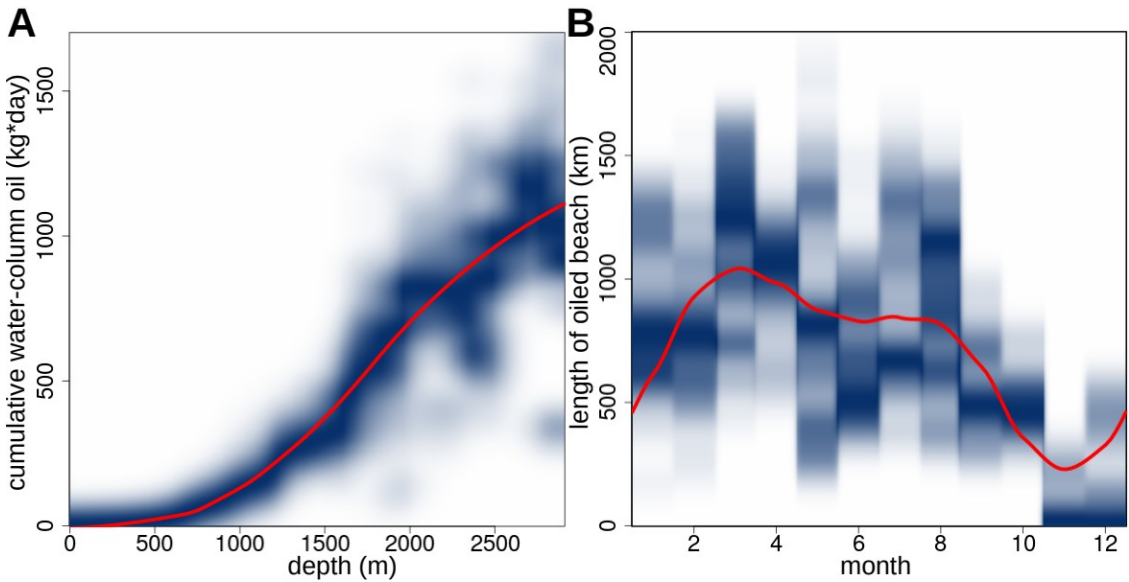

*Figure 6: Conditional distributions: (A) ocean floor depth on cumulative water column oil; (B) month on length of oiled beach. Probability density is normalized so the highest values along vertical lines are all in bark blue. Local regression curves are in red. Credit: the authors.*

The second question applies to real-time response, where some impact variables are less observable than the others. For example, the area and volume of oil slicks can be estimated using remote sensing techniques (Leifer et al., 2012), whereas crude flow rate, area of sunk oil, and mass of water-column oil are harder to measure. In this case, one may condition on the location and onset time of the blowout, along with the measured slick area and volume, to obtain a refined probabilistic estimate of the other impact variables. Symbolically, one estimates $P(Y_u|X,Y_o)$, where $Y_o$ and $Y_u$ denotes the observed and unobserved impact variables respectively.

The last question is relevant to preventative measures that reduce the likelihood and severity of future events. Ocean energy regulators may want to know the blowout scenarios that pose high risk on its coastal community and economy, and adjust the qualification standard of platform operators based on field location and time of year. This is equivalent to ask $P(X|Y_s>c)$, where $Y_s$ is some socioeconomic outcome and $c$ is the thresholds that warrant special attention. fig. 7 shows the spatiotemporal distribution of blowout scenarios causing population exposure greater than half the maximum. Exposure of coastal population and economic activity are highly correlated ($R^2=0.95$), so the result would be largely similar if considering economic exposure.

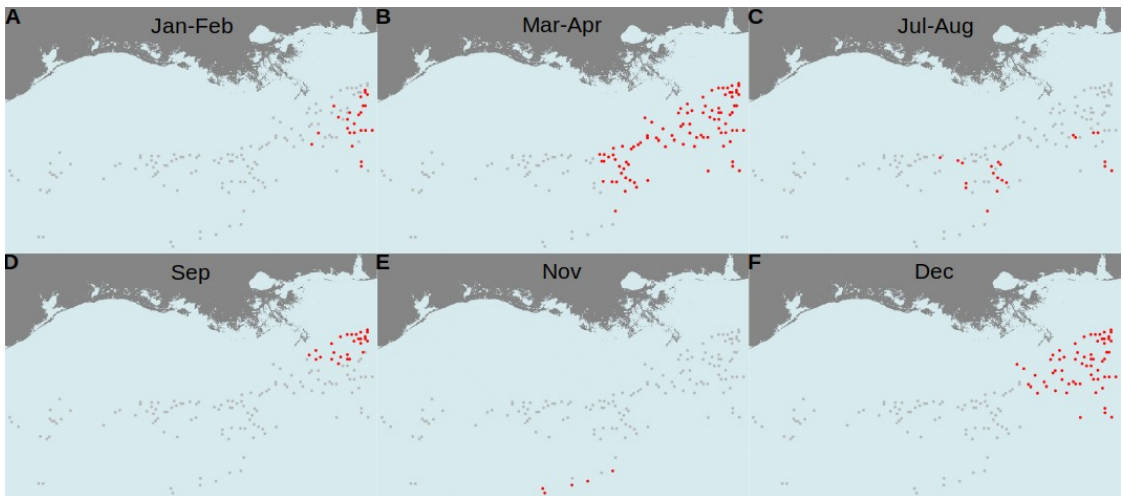

*Figure 7: Blowout scenarios causing high exposure of coastal communities. Points indicate locations of the 152 deepwater oil blocks. Scenarios exceeding half the maximum population exposure are colored red. Months with no such scenarios are omitted. Credit: the authors.*

## 6. Conclusion

This article structured environmental economic assessment in a unified framework, reviewed the current practice of uncertainty characterization within this framework, presented an alternative where manifold learning is used to build a joint probability model, and showcased its use in policy-making in the context of offshore oil spills.

This study is of course a first step toward integrated assessment of oil spills, which can be improved in several aspects. Uncertainty due to ocean current variability can be accounted for by running additional simulations that sample different ambient conditions from ocean and atmospheric models (e.g. Barker, 2011; Nelson & Grubesic, 2018). Blowout duration may be extended to capture situations when the blowout takes longer to cap. The joint distribution may be adjusted by weighting on blowout variables, based on historical observation and expert opinion. To more accurately simulate oil transport in nearshore regions, ocean models may be downscaled (Weisberg, Lianyuan, & Liu, 2017). Ocean models may also be coupled with wave models to capture Stokes drift, or supplemented with wind forcing.

Although the metrics in this case study quantify exposure rather than impact, the presented methodology is readily applicable to impact metrics once there is more understanding about vulnerability. Models connecting oil exposure and impact are limited to a subset of species and injuries to them.

The impacts of an oil spill are not summarized into one aggregate metric such as GDP; instead, a list of metrics is extracted that robustly represent the impacts to distinct groups. If a scalar summary is necessary for decision-making, the users can choose weights based on their own preferences. For example, federal, state, and local governments, environmental NGOs, and individuals may care about different aspects of an event, and thus may take different weights.

A game theory perspective may be applied to such environmental issues, because oil spills caused by one operator in the Exclusive Economic Zone of one country can spread across political boundaries and result in varied impacts at state, county, and municipality levels. A proper legal framework needs to be in place to settle accountability of all stakeholders.

## Further Reading

Hsiang et al., 2017

Houser et al., 2015

Soize & Ghanem, 2016

Duran et al., 2018

Arrow et al., 1993

## Acknowledgments

This work was supported in part by the U.S. Department of Energy, Office of Science, Office of Advanced Scientific Computing Research, Scientific Discovery through Advanced Computing (SciDAC) program through the FASTMath Institute.

This work was executed in part by NETL, AECOM and Theiss Research in support of NETL's ongoing research under the Offshore Spill Prevention Program, DE-FE1022409. This Project was conducted in part by the Department of Energy, National Energy Technology Laboratory, an agency of the United States Government, with support via contract with AECOM and Theiss Research. Neither the United States Government nor any agency thereof, nor any of their employees, nor AECOM, nor any of their employees, makes any warranty, expressed or implied, or assumes any legal liability or responsibility for the accuracy, completeness, or usefulness of any information, apparatus, product, or process disclosed, or represents that its use would not infringe privately owned rights. Reference herein to any specific commercial product, process, or service by trade name, trademark, manufacturer, or otherwise, does not necessarily constitute or imply its endorsement, recommendation, or favoring by the United States Government or any agency thereof. The views and opinions of authors expressed herein do not necessarily state or reflect those of the United States Government or any agency thereof.